\let\includefigures=\iftrue

\input harvmac

\noblackbox

\includefigures
\message{If you do not have epsf.tex (to include figures),}
\message{change the option at the top of the tex file.}
\input epsf
\def\figin{\epsfcheck\figin}\def\figins{\epsfcheck\figins}
\def\epsfcheck{\ifx\epsfbox\UnDeFiNeD
\message{(NO epsf.tex, FIGURES WILL BE IGNORED)}
\gdef\figin##1{\vskip2in}\gdef\figins##1{\hskip.5in}
\else\message{(FIGURES WILL BE INCLUDED)}%
\gdef\figin##1{##1}\gdef\figins##1{##1}\fi}
\def\DefWarn#1{}
\def\figinsert{\goodbreak\midinsert}
\def\ifig#1#2#3{\DefWarn#1\xdef#1{fig.~\the\figno}
\writedef{#1\leftbracket fig.\noexpand~\the\figno}%
\figinsert\figin{\centerline{#3}}\medskip\centerline{\vbox{
\baselineskip12pt\advance\hsize by -1truein
\noindent\footnotefont{\bf Fig.~\the\figno:} #2}}
\bigskip\endinsert\global\advance\figno by1}
\else
\def\ifig#1#2#3{\xdef#1{fig.~\the\figno}
\writedef{#1\leftbracket fig.\noexpand~\the\figno}%
\global\advance\figno by1}
\fi

\lref\RSAdS{J. Maldacena, unpublished;
E. Witten, unpublished;  S. Gubser, ``AdS/CFT
and Gravity,'' hep-th/9912001 ; S. Giddings, E. Katz and L. Randall,
``Linearized Gravity in Brane Backgrounds,''
JHEP {\bf 0003} (2000) 023, hep-th/0002091;
M. Duff and J. Liu, ``On the Equivalence of the Maldacena
and Randall-Sundrum Pictures,'' Phys. Rev. Lett. {\bf 85} (2000) 2052,
hep-th/0003237; S. Giddings and E. Katz, ``Effective Theories and
Black Hole Production in Warped Compactifications,'' hep-th/0009176;
N. Arkani-Hamed, M. Porrati and L. Randall, ``Holography and
Phenomenology,'' hep-th/0012148; R. Rattazzi and A. Zaffaroni,
``Comments on the Holographic Picture of the Randall-Sundrum Model,''
hep-th/0012248.
}
\lref\juan{J. Maldacena, ``The Large N Limit of Superconformal
Field Theories and Supergravity,'' ATMP {\bf 2} (1998) 231, hep-th/9711200.}
\lref\RSII{L. Randall and R. Sundrum, ``An Alternative to
Compactification,'' Phys. Rev. Lett. {\bf 83} (1999) 4690, hep-th/9906064.}
\lref\RSI{L. Randall and R. Sundrum, ``A Large Mass Hierarchy from
a Small Extra Dimension,'' Phys. Rev. Lett. {\bf 83} (1999) 3370,
hep-th/9905221.}
\lref\adscft {S. Gubser, I. Klebanov and A. Polyakov, ``Gauge Theory
Correlators from Noncritical String Theory,'' Phys. Lett. {\bf B428}
(1998) 105, hep-th/9802109; E. Witten, ``Anti-de Sitter Space and
Holography,'' ATMP {\bf 2} (1998) 253, hep-th/9802150.}
\lref\albion{V. Balasubramanian, P. Kraus and A. Lawrence, ``Bulk vs
Boundary Dynamics in Anti-de Sitter Space-Time,'' Phys. Rev.
{\bf D59} (1999) 046003, hep-th/9805171.}
\lref\dimo{S. Dimopoulos, ``LHC, SSC and the Universe'', Phys. Lett.
{\bf B246} (1990) 347; N. Arkani-Hamed, L.J. Hall, C. Kolda
and H. Murayama, ``A New Perspective on Cosmic Coincidence Problems",
Phys. Rev. Lett. {\bf 85} (2000) 4434, astro-ph/0005111.}
\lref\spesta{ D.N. Spergel and P.J. Steinhardt,
``Observational Evidence for Selfinteracting Cold Dark Matter",
Phys. Rev. Lett. {\bf 84} (2000) 3760, astro-ph/9909386.}
\lref\peebles{P.J.E. Peebles , ``Fluid Dark Matter", astro-ph/0002495.}
\lref\cen{Renyue Cen, ``Decaying Cold Dark Matter Model and Small-Scale
Power",  astro-ph/0005206}
\lref\joereview{J. Polchinski, ``TASI Lectures on D-branes,'' hep-th/9611050.}
\lref\dfgk{O. DeWolfe, D. Freedman, S. Gubser and A. Karch, ``Modeling
the Fifth-Dimension with Scalars and Gravity,'' Phys. Rev. {\bf D62}
(2000) 046008.}
\lref\led{N. Arkani-Hamed, S. Dimopoulos and G. Dvali, ``The
Hierarchy Problem and New Dimensions at a Millimeter,'' Phys. Lett.
{\bf B429} (1998) 263, hep-th/9803315; I. Antoniadis, N. Arkani-Hamed,
S. Dimopoulos and G. Dvali, ``New Dimensions at a Millimeter to a Fermi
and Superstrings at a TeV,''
Phys. Lett. {\bf B436} (1998) 257, hep-th/9804398.}
\lref\hw{P. Ho\v rava and E. Witten, ``Heterotic and Type I String Dynamics
from Eleven-Dimensions,'' Nucl. Phys. {\bf B460} (1996) 506, hep-th/9510209.}
\lref\HV{H. Verlinde, ``Holography and Compactification,'' Nucl. Phys.
{\bf B580} (2000) 264, hep-th/9906182.}
\lref\edunify{E. Witten, ``Strong Coupling Expansion of Calabi-Yau
Compactification,'' Nucl. Phys. {\bf B471} (1996) 135, hep-th/9602070.}
\lref\svw{S. Sethi, C. Vafa and E. Witten, ``Constraints on Low
Dimensional String Compactifications,'' Nucl. Phys. {\bf B480} (1996)
213, hep-th/9606122.}

\lref\braneworld{Recent attempts to construct such models can be found in:
G. Aldazabal, S. Franco, L.E. Ibanez, R. Rabadan and A.M. Uranga,
``D=4 Chiral String Compactifications from Intersecting Branes,''
hep-th/0011073; G. Aldazabal, S. Franco, L.E. Ibanez, R. Rabadan
and A.M. Uranga, ``Intersecting Brane Worlds,'' hep-ph/0011132;
G. Aldazabal, L.E. Ibanez, F. Quevedo and A.M. Uranga, ``D-branes
at Singularities: A Bottom Up Approach to the String Embedding of
the Standard Model,'' JHEP {\bf 0008} (2000) 002, hep-th/0005067.}

\lref\adsbreak{S. Kachru and E. Silverstein, ``4d Conformal Field
Theories and Strings on Orbifolds,'' Phys. Rev. Lett. {\bf 80}
(1998) 4855, hep-th/9802183\semi
A. Lawrence, N. Nekrasov and C. Vafa, ``On Conformal Field Theories in
Four Dimensions,'' Nucl. Phys. {\bf B533} (1998) 199, hep-th/9803015\semi
J. Distler and F. Zamora, ``Non-Supersymmetric Conformal Field Theories
from Stable Anti-de Sitter Spaces,'' Adv. Theor. Math. Ph. {\bf 2} (1999)
1405, hep-th/9810206.}

\lref\bklt{V. Balasubramanian, P. Kraus, A. Lawrence and S. Trivedi,
``Holographic probes of anti-de Sitter spacetimes,''
Phys. Rev. {\bf D59}\ (1999) 104021;
hep-th/9808017.}
\lref\bgl{V. Balasubramanian, S.B. Giddings and A. Lawrence,
``What do CFTs tell us about anti-de Sitter spacetimes?''
JHEP {\bf 9903}\ (1999) 001; hep-th/9902052.}
\lref\bdhm{T. Banks, M.R. Douglas, G.T. Horowitz and E. Martinec,
``AdS dynamics from conformal field theory,'' hep-th/9808016.}
\lref\scatt{I.R. Klebanov, ``World volume approach to
absorption by non-dilatonic branes,'' Nuc. Phys.
{\bf B496}\ (1997) 231.}
\lref\brfreed{P. Breitenlohner and D.Z. Freedman,
``Positive energy in anti-de Sitter backgrounds
and gauged extended supergravity,'' Phys. Lett.
{\bf B115}\ (1982) 197; ibid.,
``Stability in gauged extended supergravity,'' Ann. Phys. NY
{\bf 144}\ (1982) 249; L. Mezinescu and P.K. Townsend,
``Stability at a local maximum in higher-dimensional
anti-de Sitter space and applications to supergravity,''
Ann. Phys. NY {\bf 160}\ (1985) 406.}
\lref\Igor{S. Gubser, I. Klebanov, and A. Peet, ``Entropy and
Temperature of Black 3-Branes'', Phys. Rev. {\bf D54} (1996) 3915,
hep-th/9602135.}
\lref\IgorII{I. Klebanov, ``Wordvolume Approach to Absorption
by Nondilatonic Branes'', Nucl. Phys. {\bf B496} (1997) 231;
S. Gubser, I. Klebanov, A. Tseytlin, ``String Theory and Classical
Absorption by Three-Branes'', Nucl. Phys. {\bf B499} (1997) 217.}
\lref\gravbox{J. Lykken, R. Myers, J. Wang, ``Gravity in a Box'',
JHEP {\bf 0009} (2000) 009.}

\lref\TZ{
A.~A.~Tseytlin and K.~Zarembo,
``Effective potential in non-supersymmetric 
SU(N) x SU(N) gauge theory  and interactions of type 0 D3-branes,''
Phys.\ Lett.\ B {\bf 457}, 77 (1999)
hep-th/9902095.
}
\lref\typezero{
I.~R.~Klebanov and A.~A.~Tseytlin,
``A non-supersymmetric large N CFT from type 0 string theory,''
JHEP{\bf 9903}, 015 (1999)
hep-th/9901101;
I.~R.~Klebanov and A.~A.~Tseytlin,
``D-branes and dual gauge theories in type 0 strings,''
Nucl.\ Phys.\ B {\bf 546}, 155 (1999)
hep-th/9811035;
I.~R.~Klebanov and A.~A.~Tseytlin,
``Asymptotic freedom and infrared behavior in the type 0 string approach  to gauge theory,''
Nucl.\ Phys.\ B {\bf 547}, 143 (1999)
hep-th/9812089;
I.~R.~Klebanov, N.~A.~Nekrasov and S.~L.~Shatashvili,
``An orbifold of type 0B strings and non-supersymmetric gauge theories,''
Nucl.\ Phys.\ B {\bf 591}, 26 (2000)
hep-th/9909109.
}
\lref\ksorb{
S.~Kachru and E.~Silverstein,
``4d conformal theories and strings on orbifolds,''
Phys.\ Rev.\ Lett.\ {\bf 80}, 4855 (1998)
hep-th/9802183.
}
\lref\otherorb{
A.~E.~Lawrence, N.~Nekrasov and C.~Vafa,
``On conformal field theories in four dimensions,''
Nucl.\ Phys.\ B {\bf 533}, 199 (1998)
hep-th/9803015;
M.~Berkooz,
``A supergravity dual of a (1,0) field theory in six dimensions,''
Phys.\ Lett.\ B {\bf 437}, 315 (1998)
hep-th/9802195.
}
\lref\kks{
S.~Kachru, J.~Kumar and E.~Silverstein,
``Vacuum energy cancellation in a non-supersymmetric string,''
Phys.\ Rev.\ D {\bf 59}, 106004 (1999)
hep-th/9807076.
}
\lref\dougmoore{
M.~R.~Douglas and G.~Moore,
``D-branes, Quivers, and ALE Instantons,''
hep-th/9603167.
}
\lref\Igortach{
I.~R.~Klebanov,
``Tachyon stabilization in the AdS/CFT correspondence,''
Phys.\ Lett.\ B {\bf 466}, 166 (1999)
hep-th/9906220.
}
\lref\joetach{
S.~P.~de Alwis, J.~Polchinski and R.~Schimmrigk,
``Heterotic Strings With Tree Level Cosmological Constant,''
Phys.\ Lett.\ B {\bf 218}, 449 (1989).
}
\lref\anttach{
I. Antoniadis and C. Kounnas, "Superstring Phase Transition at High Temperature", Phys. Lett. {\bf B261} (1991) 369.}
\lref\polyakov{
A.~M.~Polyakov,
``The wall of the cave,''
Int.\ J.\ Mod.\ Phys.\ A {\bf 14}, 645 (1999)
hep-th/9809057;
A.~M.~Polyakov,
``String theory and quark confinement,''
Nucl.\ Phys.\ Proc.\ Suppl.\ {\bf 68}, 1 (1998)
hep-th/9711002.
}
\lref\blumdienes{
J.~D.~Blum and K.~R.~Dienes,
``Strong/weak coupling duality relations for non-supersymmetric string  theories,''
Nucl.\ Phys.\ B {\bf 516}, 83 (1998)
hep-th/9707160.
}
\lref\minahan{
J.~A.~Minahan,
``Asymptotic freedom and confinement from type 0 string theory,''
JHEP{\bf 9904}, 007 (1999)
hep-th/9902074.
}
\lref\kutstud{
V.~Niarchos,
``Density of states and tachyons in open and closed string theory,''
hep-th/0010154.
}
\lref\harvey{
J.~A.~Harvey,
``String duality and non-supersymmetric strings,''
Phys.\ Rev.\ D {\bf 59}, 026002 (1999)
hep-th/9807213.
}
\lref\kkstach{
S.~Kachru, J.~Kumar and E.~Silverstein,
``Orientifolds, RG flows, and closed string tachyons,''
Class.\ Quant.\ Grav.\ {\bf 17}, 1139 (2000)
hep-th/9907038.
}
\lref\bosM{
G.~T.~Horowitz and L.~Susskind,
``Bosonic M theory,''
hep-th/0012037.
}
\lref\berggab{
O.~Bergman and M.~R.~Gaberdiel,
``A non-supersymmetric open-string theory and S-duality,''
Nucl.\ Phys.\ B {\bf 499}, 183 (1997)
hep-th/9701137.
}
\lref\opentach{
E.~Witten,
``Some computations in background independent off-shell string theory,''
Phys.\ Rev.\ D {\bf 47}, 3405 (1993)
hep-th/9210065;
J.~A.~Harvey, D.~Kutasov and E.~J.~Martinec,
``On the relevance of tachyons,''
hep-th/0003101;
D.~Kutasov, M.~Marino and G.~Moore,
``Some exact results on tachyon condensation in string field theory,''
JHEP{\bf 0010}, 045 (2000)
hep-th/0009148;
A.~A.~Gerasimov and S.~L.~Shatashvili,
``On exact tachyon potential in open string field theory,''
JHEP{\bf 0010}, 034 (2000)
hep-th/0009103;
S.~L.~Shatashvili,
``On the problems with background independence in string theory,''
hep-th/9311177;
S.~L.~Shatashvili,
``Comment on the background independent open string theory,''
Phys.\ Lett.\ B {\bf 311}, 83 (1993)
hep-th/9303143.
}
\lref\bkv{
M.~Bershadsky, Z.~Kakushadze and C.~Vafa,
``String expansion as large N expansion of gauge theories,''
Nucl.\ Phys.\ B {\bf 523}, 59 (1998)
hep-th/9803076.
}
\lref\vafaquant{
C.~Vafa,
``Quantum Symmetries Of String Vacua,''
Mod.\ Phys.\ Lett.\ A {\bf 4} (1989) 1615.
}
\lref\ksdual{ S.~Kachru and E.~Silverstein,
``Self-dual nonsupersymmetric type II string compactifications,''
JHEP{\bf 9811}, 001 (1998)
hep-th/9808056.
}
\lref\kutsei{
D.~Kutasov and N.~Seiberg,
``Number of degrees of freedom, density of states and tachyons in string theory and CFT,''
Nucl.\ Phys.\ B {\bf 358}, 600 (1991).
}
\lref\polstrass{
J.~Polchinski and M.~J.~Strassler,
``The string dual of a confining four-dimensional gauge theory,''
hep-th/0003136.
}
\lref\garousi{
M.~R.~Garousi,
``String scattering from D-branes in type 0 theories,''
Nucl.\ Phys.\ B {\bf 550}, 225 (1999)
hep-th/9901085.
}
\lref\BF{
P.~Breitenlohner and D.~Z.~Freedman,
``Positive Energy In Anti-De Sitter Backgrounds And Gauged Extended Supergravity,''
Phys.\ Lett.\ B {\bf 115}, 197 (1982).
}
\lref\DT{
M.~R.~Douglas and B.~Fiol,
``D-branes and discrete torsion. II,''
hep-th/9903031.
M.~R.~Douglas,
``D-branes and discrete torsion,''
hep-th/9807235.
}
\lref\berleigh{
D.~Berenstein and R.~G.~Leigh,
``Discrete torsion, AdS/CFT and duality,''
JHEP{\bf 0001}, 038 (2000)
hep-th/0001055.
}
\lref\CW{
S.~Coleman and E.~Weinberg,
``Radiative Corrections As The Origin Of Spontaneous Symmetry Breaking,''
Phys.\ Rev.\ D {\bf 7} (1973) 1888.
}
\lref\tuninst{
E.~Witten,
``Instability Of The Kaluza-Klein Vacuum,''
Nucl.\ Phys.\ B {\bf 195}, 481 (1982);
M.~Fabinger and P.~Horava,
``Casimir effect between world-branes in heterotic M-theory,''
Nucl.\ Phys.\ B {\bf 580}, 243 (2000)
hep-th/0002073;
}
\lref\gary{
F.~Dowker, J.~P.~Gauntlett, G.~W.~Gibbons and G.~T.~Horowitz,
``The Decay of magnetic fields in Kaluza-Klein theory,''
Phys.\ Rev.\ D {\bf 52}, 6929 (1995)
hep-th/9507143.
}
\lref\maldstrom{ J.~Maldacena, J.~Michelson and A.~Strominger,
``Anti-de Sitter fragmentation,''
JHEP{\bf 9902}, 011 (1999)
hep-th/9812073.
}
\lref\gutperle{
M.~S.~Costa and M.~Gutperle,
``The Kaluza-Klein Melvin solution in M-theory,''
hep-th/0012072.
}
\lref\yun{
E.~Silverstein and Y.~S.~Song,
``On the critical behavior of D1-brane theories,''
JHEP{\bf 0003}, 029 (2000)
hep-th/9912244.}
\lref\ofer{O. Aharony, M. Berkooz, and E. Silverstein, work in progress.}
\lref\Ittach{
C. Angelantonj and A. Armoni, ``RG flow, Wilson Loops, and the Dilaton 
Tadpole'', Phys. Lett. B {\bf 482} 329 (2000);
``Non-Tachyonic type 0B Orientifolds, Non-supersymmetric Gauge Theories, and
Cosmological RG Flow'', Nucl. Phys. B {\bf 578} 239 (2000). 
}
\lref\othercalcs{
C. Csaki, W. Skiba, and J. Terning, ``Beta Functions of Orbifold
Theories and the Hierarchy Problem'',  Phys. Rev. D {\bf 61},
025019 (2000); 
P. H. Frampton and W. F. Shively, ``Conformal N=0 d=4 Gauge Theories
from AdS/CFT Superstring Duality?'', Phys. Lett. B {\bf 454} 49 (1999). 
}
\lref\typezorb{N. Nekrasov and S. Shatashvili, ``On Non-Supersymmetric
CFT in Four Dimensions'', Phys. Rept. {\bf 320}, 127 (1999);
E. Silverstein, unpublished.  
}
\lref\hirosijuan{
J.~Maldacena and H.~Ooguri,
``Strings in AdS(3) and SL(2,R) WZW model. I,''
hep-th/0001053.
}
\lref\bigazzi{
F. Bigazzi, ``RG flows toward IR isolated fixed points:  Some
type 0 samples'', hep-th/0101232.
}


\def\frac#1#2{{#1 \over #2}}

\Title{\vbox{\baselineskip12pt\hbox{hep-th/0103220}
\hbox{SLAC-PUB-8797}}}
{\vbox{
\centerline{Closed String Tachyons, AdS/CFT,}
\bigskip
\centerline{and Large N QCD}
}}
\medskip
\centerline{Allan Adams and Eva Silverstein}
\bigskip
\centerline{Institute for Theoretical Physics; University of California;
Santa Barbara, CA 93106}
\centerline{Department of Physics and SLAC; Stanford University; Stanford,
CA 94305/94309\foot{allan, 
evas@slac.stanford.edu}}
\bigskip
\noindent

We find that tachyonic orbifold examples 
of AdS/CFT have corresponding instabilities at small radius,
and can decay to more generic gauge theories.
We do this by computing a destabilizing 
Coleman-Weinberg effective potential for
twisted operators of 
the corresponding quiver gauge theories, generalizing calculations
of Tseytlin and Zarembo and interpreting them in terms of the large-N
behavior of twisted-sector modes.  The dynamically generated
potential involves double-trace operators, which affect large-N correlators
involving twisted fields but not those involving only untwisted fields, in
line with large-N inheritance arguments.  
We point out a simple reason that no such 
small radius instability exists in gauge
theories arising from freely acting orbifolds, 
which are tachyon-free at large radius.
When an instability is present, twisted gauge theory operators with the quantum
numbers of the
large-radius tachyons aquire VEVs, leaving a gauge theory with fewer
degrees of freedom in the infrared, analogous to but less extreme
than ``decays to nothing'' studied in other
systems with broken supersymmetry.  In some cases one is left
with pure 
glue QCD plus decoupled matter and U(1) factors in the IR, which
we thus conjecture is described by the corresponding
(possibly strongly coupled) endpoint of tachyon
condensation in the M/String-theory dual.

\smallskip

\Date{March 2001}


\newsec{Introduction and Summary}

Orbifold examples \ksorb\otherorb\typezero\ provide one of the simplest testing
grounds and applications of AdS/CFT duality \juan\adscft.  With less than
maximal SUSY, the physics at low energies is less constrained, and new elements
of the AdS/CFT dictionary emerge. One such element is the relation between the
gravity-side cosmological term (dilaton potential) which typically gets
generated in the absence of SUSY, and the finite-N beta functions and
dimension spectrum of the gauge theory \ksorb\kks\Ittach.

Another element that arises upon breaking supersymmetry is the possibility of
a stringy tachyon in the twisted sector on the gravity side.
This typically (in fact in all cases known to the authors) happens at large
AdS radius when the orbifold is symmetric and fixes some or all of the points
on the $S^q$ component of the gravity background.\foot{A tachyon indeed
appears in the twisted sector of the non-SUSY model studied in detail in
\ksorb; the statement otherwise was in error.} Freely-acting orbifolds on
the $S^q$ have no twisted-sector tachyon at large radius, since the
twisted-sector states must
wind around the sphere and are therefore very heavy. In this paper, we
find an interesting pattern in the corresponding instability structure
of the small radius limit of these theories by investigating the dynamics of
twisted operators in the appropriate weakly-coupled dual quiver gauge theories
\dougmoore.

Building on work of Tseytlin and Zarembo \TZ, we study radiative corrections
in $IIB/(-1)^F$ (Type 0) on $AdS_5\times S^5$ 
and more general non-supersymmetric non-freely-acting orbifolds
by $\Gamma=Z_n$, identifying a Coleman-Weinberg effective 
potential \CW\ which leads to
growth of the VEVs of certain twisted 
operators quadratic in the gauge theory scalars. This
instability along a (partial) Coulomb branch of the quiver gauge theory describes
D-branes splitting into fractional branes at the fixed point
locus of the orbifold. This one-loop contribution to the effective potential
involves double-trace operators which affect correlators involving twisted
operators at leading order in $1/N^2$ (corresponding to genus
zero on the gravity side).  It is the leading non-conformal effect at
small 't Hooft coupling and introduces non-conformal correlators for twisted
operators into the theory at large $N$. 

This raises an interesting puzzle some of whose potential
resolutions we discuss in \S3.2.  It has to do with the
issue of whether and how the conformal symmetry of the gauge
theory (or equivalently the $SO(4,2)$ isometry group of the
$AdS_5\times S^5/Z_n$) should be preserved in correlation functions
involving twisted operators in the QFT dual to the 
standard gravity-side orbifold construction of the
$AdS_5\times S^5$ sigma model.  As we will discuss in \S2,
there is a space of possible orbifold quantum field theories
arising from the space of renormalization group
trajectories consistent with the quantum symmetry
and inheritance of untwisted operator correlation
functions from the ${\cal N}=4$ theory.  
This suggests a corresponding space of dual
gravity-side orbifold string backgrounds, generic elements
of which may generalize
the standard construction.                

The one-loop calculation is reliable in this context
for a large but finite range of nonzero VEVs for the scalar fields,
including values near a
local minimum of the effective potential. We find that a renormalization-group
improved perturbation theory analysis using one-loop beta functions
does not lead to any additional control for the calculation of the 
potential near the origin of the Coulomb branch. From this we conclude that
the instability indicated by this potential is present but,
depending on the form of the potential near the origin of the Coulomb branch,
may be a global effect accessible only by tunneling.

The resulting instability, present in the
large-N limit, drives the erstwhile tachyon (or at least a twisted
state with the same quantum numbers as a large-radius tachyon) to condense.
It is interesting to note that whatever
the behavior near the origin, the instability is
consistent with the reality of the operator dimensions observed in
\Igortach, though not with an uncorrected extrapolation of the
Breitenlohner-Freedman bound to the regime of large curvatures
and large RR field strengths.

For a freely-acting orbifold, the quiver gauge theory has no Coulomb
branch, as the D-branes are at the same codimension as the orbifold fixed point
locus in spacetime and have no directions in which to split up
into separated fractional branes.  
One correspondingly finds no instability in the effective
potential for twisted operators; we therefore conjecture that the tachyon-freedom
of this type of model persists to small radius.  This constitutes an interesting
prediction for the RR string theory in this limit, and a satisfying class of
examples to contrast with the tachyonic models.

We also study a case with non-free orbifold action and discrete torsion
which projects out the lowest-lying tachyon, leaving tachyonic modes with 
angular momentum along the sphere.  At weak coupling we find no one-loop
instability.  However there is a Coulomb branch (again describing
fractional D-branes) in this
case, and we find no symmetries preventing 
instabilities from getting generated at higher order. 

We then consider in more detail the effects of condensing the twisted modes in the unstable
(non-freely-acting) cases.  On the gravity side at large radius, condensing a
tachyon is expected to drive the theory to a subcritical dimension (with
compensating gradients for the dilaton and other gravity fields), since the
zero-momentum tachyon is a relevant operator on the string worldsheet
\joetach\anttach\kkstach.  One way in which this can happen is to lose 
and/or deform the dimensions
corresponding to the $S^5/\Gamma$ and the radial direction in 
$AdS_5$.  This suggests losing the scalars and perhaps
non-chiral fermions on the gauge theory side, since these matter fields
have quantum numbers associated to the sphere. Similarly, non-perturbative
instabilities have been argued to drive the theory to an endpoint with a loss of
degrees of freedom (decays to ``nothing") \tuninst.

We find on the field theory side that the Coleman-Weinberg potential indeed drives the
theory toward one with fewer degrees of freedom at
least in the IR:  in some cases, pure
glue $SU(N)$ QCD plus some decoupled matter and U(1) factors. 
This can be seen algebraically or from a simple geometric picture of 
the low-energy/near horizon limit of 
symmetrically distributed fractional D-branes.  
We conjecture that this
pure glue QCD theory is described by the endpoint of tachyon condensation
in the dual gravity theory, at least at long distance on the QFT side.  As just
discussed, this is likely to be a $D<10$ theory on the gravity side (which may
or may not be a perturbative string theory, depending on the effective string
coupling that emerges in the subcritical theory when the tachyon has become large
and mixed with the dilaton and other string fields).  This result is suggestive of
Polyakov's proposal for QCD as a noncritical string \polyakov; our analysis
suggests that if realized it could be connected to ordinary AdS/CFT by tachyon
condensation. \foot{ This possibility was suggested earlier in general terms
by Minahan at the end of \minahan.} More complicated models involving both open
and closed strings on the gravity side may leave surviving quark flavors as
well, but we leave this goal of getting full QCD for future work.

Of course there are other avenues toward the String/M dual of QCD, arising from the
low-energy limit of relevant perturbations of the ${\cal N}=4$ theory as in
for example \polstrass.  Our connection to non-SUSY AdS/CFT examples via tachyon
condensation appears complementary to these.  It involves a 
dynamical mechanism for eliminating the
extra matter on the gauge theory side, but the
starting point is a non-supersymmetric system less well-understood than the
${\cal N}=4$ theory.  As with many applications of AdS/CFT, it can be taken 
as a lesson about the gravity side: the gauge theory remaining after condensation
of the twisted operators gives a dual answer to the question of what happens
to the closed string theory after tachyon condensation.

There have been interesting discussions on closed string tachyons in D-brane and/or
AdS/CFT systems in \typezero\Igortach\yun\kutstud\bigazzi.  
The role of tachyons,
tachyon condensation, and/or tachyon-freedom in other 
conjectured non-supersymmetric closed
string dualities has been studied in other contexts previously for example in
\blumdienes\berggab\kks\harvey\ksdual\kkstach\bosM.  Calculations making use of
the relevance of open-string tachyon vertex operators appeared in for example
\opentach.  It would be very interesting if these techniques could be transported
back to closed-string theory to investigate further the hints of QCD emerging
in the systems we study here.  Indeed, there is some resemblance of the
Coleman-Weinberg potentials we discuss here and formulas for an (open string)
tachyon potential in \opentach.

It would be interesting to study potential relations of our results on
instabilities on the gauge theory side to the kinds of gravitational instabilities
studied in \tuninst\gary\gutperle\ and in particular in the AdS/CFT context
in \maldstrom.  Aside from a few comments, we leave this for future work, and focus
here on the field theory side at weak coupling.

The paper is organized as follows.  In \S2, we calculate the one-loop Coleman-Weinberg
potential for a large class of $Z_n$ orbifold models with fixed
points, exhibiting a nontrivial regime of validity of the one-loop
analysis out on the classical Coulomb branch.  
In \S3, we find two simple minima of the potential within this regime of validity
at which the low-energy gauge theory contains pure glue QCD sectors.  We
study the N-dependence of our results from \S2\ and establish that 
at large N the
dynamically generated potential affects correlators involving twisted
operators, and does not affect untwisted correlators.  
We discuss a puzzle this raises and several
possibilities which may lead to its resolution, and review the evidence
we have gathered for the persistence of the tachyon to small radius in
these examples.  In \S4\ we turn to the case of freely-acting orbifolds
(where there is no tachyon at large radius) 
and show that the branch along which such an instability would
arise is absent in this case.  In \S5\
we analyze an orbifold with discrete torsion which has no Coleman-Weinberg
instability at one loop but does have a Coulomb branch along which one
could emerge at higher orders.  Finally in \S6\ we discuss future directions.

\newsec{$Z_n$ Orbifolds with Fixed Points and Effective Potential}

Given the duality between ${ \cal N}=4$ SYM and IIB string theory on $AdS_5\times
S^5$, we can obtain new dual pairs by orbifolding both sides of this
``parent'' duality by a discrete group 
$\Gamma\subset SO(6)$ \ksorb\otherorb.  The gravity side is Type IIB string theory
on $AdS_5\times S^5/\Gamma$, and the QFT side is a quiver gauge theory
\dougmoore\ obtained by taking the low-energy limit of the worldvolume theory
on D3-branes at the corresponding $C^3/\Gamma$ orbifold singularity. \foot{We
are working with orbifolds of the maximally supersymmetric version
of the duality. The duality being a statement of equivalence between two
descriptions of one and the same system, the orbifold (including its instabilities)
tautologically exists on one side if it exists on the other.  
Note that this equivalence is between the quantum theories, which introduces
subtleties having to do with the manifestation on the gravity
side of the choice of renormalization condition
in the gauge theory side as we will discuss shortly.} 
On the gravity side, correlations of untwisted operators are
inherited at genus zero.  The corresponding planar diagrams, in particular those
corresponding to the beta functions \ksorb, of the quiver gauge theories are
inherited from those of the parent ${\cal N}=4$ theory \bkv.\foot{ It is worth
emphasizing that we here take the regular representation for the action $\gamma_{ab}$
of the orbifold group on the Chan-Paton indices, one where $tr\gamma =0$.
With any other choice of action, the D-branes are a source for the twisted-sector
tachyon in the asymptotically flat region away from the core
of the D-branes, which does not decouple from the near-horizon
low energy D-brane theory 
\kutstud.}

Twisted states in the orbifold string theory correpond to ``twisted operators'' of
the orbifold gauge theory, which are gauge-invariant operators that do not
descend from gauge-invariant operators of the parent ${\cal N}=4$ theory.
On the gravity side, non-freely-acting geometrical orbifolds have tachyons in the
twisted sector at least at large radius, while freely-acting orbifolds do not have
tachyons, since twisted-sector states are very massive; we
will return to them in \S4.  The effective potential for twisted operators is
 not constrained by the ${\cal N}=4$ theory, 
so we should expect an effective potential for twisted operators at
order $N^2$, whereas the effective potential for untwisted operators should appear only at  order $N^0$;
we will see this explicitly below.

We will find it instructive to study several different cases.  In this section
we will focus on non-freely-acting orbifolds, which fix some locus on the
$S^5$.  The most extreme case of this is the ``Type 0'' theory \typezero, which
is an orbifold by $(-1)^F$ which acts only on the spinors and thus fixes the
entire spacetime \typezorb.  The corresponding quiver theory has gauge group $SU(N)^2$ with
six real adjoint scalars $X^i_1=({\bf N^2-1, 1})$, $X^i_2=({\bf 1, N^2-1})$,
$i=1,\dots, 6$, four Weyl fermions $\chi^\alpha$ in the bifundamental
representation $(\bf {N, \bar N})$, and four Weyl fermions $\psi^\alpha$ in the $(\bf
{\bar N, N})$.

These representations arise from projections of a parent $SU(2N)$ ${\cal N}=4$ theory.  In a convenient basis, the gauge fields and scalars sit in diagonal
blocks of this $SU(2N)$ theory and the fermions 
sit in the off-diagonal blocks.  The tree-level interactions 
of the orbifold theory are those of the $SU(2N)$
${\cal N}=4$ theory which involve fields which survive 
the orbifold projection.  In addition to the minimal gauge couplings, one has quartic scalar
interactions of the $X^i_1$ which are identical 
to those of an $SU(N)$ ${\cal N}=4$ theory and 
likewise for $X^i_2$.  The bifundamental fermions mix the two
$SU(N)$ gauge groups via tree-level Yukawa interactions of the 
form\foot{In this section we henceforth assume canonical normalization for the kinetic terms with no
factors of $1/g_{YM}^2$ in front of the action.}
\eqn\typezyuk{
g_{YM}\biggl[
tr(\bar\psi X_1 \psi) + tr(\bar\chi\chi X_1)
+tr(\bar\chi X_2 \chi) + tr(\bar\psi\psi X_2)
\biggr]
}
The orbifold on the gravity side has a ``quantum symmetry'' 
$\Gamma$, \vafaquant, under which twisted states transform non-trivially. In the Type 0 case, this
is a $Z_2$ symmetry which is manifested on the gauge theory side by a symmetry exchanging the two $SU(N)$ factors in the gauge group (and correspondingly
exchanging $X^i_1\leftrightarrow X^i_2$ and $\psi\leftrightarrow\chi$).  The 
lowest-dimension twisted operators in our theory are thus $tr X_1^2 -tr X_2^2$
and $tr X_1^iX_1^j-tr X_2^iX_2^j$,
which have dimension two at leading order. In the full D-brane system, 
these operators couple to two derivatives of the tachyon field in the directions
transverse to the D-branes \garousi, while the 
tachyon itself couples to the dimension 
four operator $tr (F_1^2+DX_1DX_1-F_2^2-DX_2DX_2)$ \garousi\Igortach.

The interactions of untwisted states are inherited at genus zero, but 
those of twisted states are constrained a priori only by the quantum symmetry.  Our main
interest will be contributions to the scalar potential generated by quantum corrections 
to the gauge theory at leading order in the $1/N^2$ expansion. We will discuss the
calculation of the effective potential in non-freely-acting $Z_n$ 
orbifolds in the remainder of this section, and then interpret the results in terms of
tachyons and explore the instability structure on the gauge theory side in \S3.

In \TZ, Tseytlin and Zarembo calculated the bosonic potential energy lifting the classical moduli space of the Type 0 theory at one-loop order in the gauge
theory.  Because of the quartic scalar interactions, the classical moduli space of this theory is parameterized by the eigenvalues of commuting matrices
$\langle X^i_1\rangle \equiv diag(x^{i,1}_1, \dots, x^{i,N}_1)$ and $\langle X^i_2\rangle \equiv diag(x^{i,1}_2, \dots, x^{i,N}_2)$ with $\Sigma_a
x_1^{i,a}=0=\Sigma_a x_2^{i,a}$.  Going to a generic point on this moduli space and integrating out massive particles, one obtains a simple expression for the
vacuum energy as a function of $x^{i, a}_1$ and $x^{i, a}_2$ \TZ :
\eqn\tzpot{\eqalign{
& V_{eff} \sim {g_{YM}^4\over{8\pi^2}}
\Sigma_{a, b=1}^N \biggl[ |x^a_1-x^b_1|^4
log{{|x^a_1-x^b_1|^2}\over{\tilde M^2}} + |x^a_2-x^b_2|^4
log{{|x^a_2-x^b_2|^2}\over{\tilde M^2}}\cr
& -2|x^a_1-x^b_2|^4
log{{|x^a_1-x^b_2|^2}\over{\tilde M^2}}\biggr]\cr
}
}

where $|x|^2\equiv x^i x^i\equiv \vec x^2$ 
and $\tilde M$ is related to a subtraction point to be discussed shortly.  The first two terms in \tzpot\ 
arise from integrating
out the off-diagonal entries in $(\vec X_1)_{ab}, (A_1^\mu)_{ab}$ and $(\vec X_2)_{ab}, (A_2^\mu)_{ab}$, 
which have masses $g_{YM}|x^{a}_1-x^{b}_1|$ and
$g_{YM}|x^{a}_2-x^{b}_2|$ respectively.  The last arises from integrating out the fermions $(\psi)_{ab}$ 
and $(\chi)_{ab}$ in the bifundamental $({\bf N,
\bar N})\oplus({\bf \bar N, N})$, whose masses are $g_{YM}|x^a_1-x^b_2|$. This expression can be understood (and later generalized) as follows.  Integrating
out a particle of mass $m^2$ leads to a contribution
\eqn\basiccont{
(-1)^{F}\int d^4 p ~~ log(p^2+m^2)\sim
(-1)^{F}\int_0^\infty {{dt}\over t^3}e^{-tm^2}
}
where $F$ is the spacetime fermion number and where we are ignoring coefficients of order one. This expression has quadratic and logarithmic divergences (as
well as a quartic divergence in the vacuum energy to which the field theory is insensitive).  We therefore require counterterms; following the analysis of \CW\
one obtains an expression of the form
\eqn\renpot{\eqalign{
& V_{eff} =
\Sigma_{a, b=1}^N \biggl[ |x^a_1-x^b_1|^4
\bigl(\lambda_{11}^{ab}+(Ag_{YM}^4+B(\lambda_{11}^{ab})^2)
log{{|x^a_1-x^b_1|^2}\over{M^2e^{25/6}}}\bigr)\cr
& +|x^a_2-x^b_2|^4
\bigl(\lambda_{22}^{ab}+(Ag_{YM}^4+B(\lambda_{22}^{ab})^2)
log{{|x^a_2-x^b_2|^2}\over{M^2e^{25/6}}}\bigr)\cr
&-2|x^a_1-x^b_2|^4
\bigl(\lambda_{12}^{ab}+(Ag_{YM}^4+B(\lambda_{12}^{ab})^2)
log{{|x^a_1-x^b_2|^2}\over{M^2e^{25/6}}}\bigr)\biggr]\cr
}
}
where $A$ and $B$ are constants of order 1 and $g_{YM}$ and $\lambda_{ij}^{ab}$ are renormalized couplings.  This potential includes the 1-loop contributions
plus counterterms chosen to satisfy the renormalization conditions
\eqn\rencond{
{{d^4V_{eff}}\over{dx_{ij}^{ab ~~ 4}}}|_{x_{ij}^{ab}=M}=\lambda_{ij}^{ab}
}
where $x_{ij}^{ab}=x_i^a-x_j^b$.  The coupling constants determined at one value of the subtraction point $M$ are related to those at a different point by the
renormalization group.  A choice of renormalization group trajectory is a choice of field theory and presumably corresponds to a choice of what the precise
configuration of dual gravity-side string fields is.  At small radius, we do not have an independent handle on the gravity side, so we will simply consider the
whole set of possible trajectories consistent with the symmetries and inheritance properties of the orbifold.  This issue will be discussed further in \S3.2.  As discussed in \TZ, there are planar
contributions proportional to $N$ in the individual terms in the effective potential that must cancel in the orbifold theory by inheritance.  This plus the
quantum symmetry leads to the simplification that in the quantum field theory dual to the orbifold background, we should have $\lambda_{ij}^{ab}\equiv\lambda$.

As discussed in \TZ\ (and as will be generalized and studied further in \S3) there is an unstable direction in the potential in which
$x_1^1=\rho=-x_1^2$, with all other $x_i^a=0$. Plugging this into \renpot, one finds
\eqn\potrho{
V_{eff}\sim \rho^4\biggl[\lambda+(Ag_{YM}^4+B\lambda^2)
log {{2^{8/3}\rho^2}\over{M^2e^{25/6}}}\biggr]
}
Let us now renormalize at a subtraction point of order the VEV of $\rho$, e.g. $M=2^{4/3}\langle\rho\rangle$, where
\eqn\mincond{
{{d V_{eff}}\over{d\rho}}|_{\rho=\langle\rho\rangle}=0.
}
Imposing this condition, one obtains in the theory expanded about the minimum of the potential a relation between $\lambda$ and $g_{YM}$ as in \CW\ of the form
\eqn\relcoup{
\lambda = Cg_{YM}^4
}
where $C$ is a constant of order 1.  So the renormalized quartic scalar coupling along the Coulomb branch is of order $g_{YM}^4$, as befits a contribution at
one-loop order in perturbation theory.  Plugging this back into \renpot\ and defining
\eqn\massdef{
\tilde M^2\equiv e^{{{25}\over 6}-{C\over A}}M^2,
}
we recover the result \tzpot.

The one-loop result is reliable where the logarithms in \renpot\ are not big enough to compensate the small couplings $\lambda, g_{YM}$ and make different
orders in perturbation theory commensurate.  
%
%
As in the original analysis of the massless Abelian Higgs model in \CW,
our result is reliable near minima
of the potential but not at the origin $x_{ij}^{ab}\to 0$
or in the asymptotic region $x_{ij}^{ab}\to \infty$, 
since there the logarithms are large.
In some theories, expressing the effective potential in terms of the running coupling (the solution of the Callan-Symanzik
equations)  results in a weakly coupled description
for a larger range of $x$.  As demonstrated below,
this is not the case at one-loop order at large N in our theories, which
have a somewhat remarkable RG structure due to the vanishing
of untwisted beta functions at large N.\foot{We thank D. Gross for
interesting discussions on this.}

The $1$-loop $\beta$ functions are easily computed.  
For the Type 0 theory and the other quiver theories we are
about to analyze, Large-N inheretance
ensures that the gauge and yukawa couplings have vanishing $\beta$ functions at leading order in $1\over{N}$\bkv.

The $\beta$ function for the quartic scalar coupling
$(tr X_1^2- tr X_2^2)^2$ can be calculated 
directly from our calculation of the renormalized
coupling:
\eqn\betatwisted{
\beta_{\lambda} \sim \lambda^2 + g^4_{YM}
}
the resulting RG equations are solved by
\eqn\lambrun{
\lambda = g^2_{YM}~Tan(g^2_{YM}ln{{\rho^2}\over{M^2}} + g^2_o)
}

As $\rho$ gets either very large or small compared to 
$M^2~e^{\pi\over{g^2_{YM}}}$, this solution becomes strongly
coupled and untrustworthy.  So in these theories at one-loop
order at large N, RG improvement does not help.
While we can trust our one-loop effective potential near its local minimum, we cannot trust the dynamics near the
origin, or at large values of the scalar VEV.
Since the gauge coupling is protected from developing a $\beta$ function at large $N$ by inheritance, the main effect of higher loops will be
to add higher monomials in $\lambda$, whose effects will depend strongly on their signs.   
We will leave this much more involved two-loop 
calculation to future work, and in this paper content ourselves with having identified at least a global instability.
This leaves open the possibility that the region near $x\sim M$ 
could only be accessible via tunneling from the region of the origin.  We will comment
further on this in \S3.%

More generically one can consider locally-free orbifold actions. One example we
will study in detail is a $C/Z_3$ orbifold, under which a single
complex plane with coordinate
$z^1$ is rotated by $\alpha\equiv e^{2\pi/3}$: $z^1\to \alpha^2 z^1$,
$z^{2,3}\to z^{2,3}$.
This acts by a phase
$\alpha^{\pm 1}\equiv e^{\pm 2\pi/3}$ on all the spacetime spinors, and
so projects out all the massless gravitinos.  The quiver theory
in this case has gauge group $SU(N)^3$.  The
matter content consists of four real scalars in
the adjoint:
\eqn\zthreematI{\eqalign{
& X^i_1 ~~~ ({\bf N^2-1, 1, 1})\cr
& X^i_2 ~~~ ({\bf 1, N^2-1, 1})\cr
& X^i_3 ~~~ ({\bf 1, 1, N^2-1}),\cr
}
}
for $i=1, \dots, 4$;
one complex scalar in the bifundamental representations:
\eqn\zthreematII{\eqalign{
& U ~~~ ({\bf N, \bar N, 1})\cr
& V ~~~ ({\bf 1, N, \bar N})\cr
& W ~~~ ({\bf \bar N, 1, N}),\cr
}
}
and four Weyl fermions in the bifundamental representations:
\eqn\zthreematIII{\eqalign{
& \chi^\alpha_U ~~~ ({\bf N, \bar N, 1})\cr
& \chi^\alpha_V ~~~ ({\bf 1, N, \bar N})\cr
& \chi^\alpha_W ~~~ ({\bf \bar N, 1, N}),\cr
}
}
for $\alpha=1,\dots, 4$.
The interactions in this case are inherited from an $SU(3N)$
${\cal N}=4$ theory.  In a convenient basis,
the gauge bosons and adjoint scalars sit in diagonal $N\times N$
blocks of the adjoint matrices of the parent theory, and the
bifundamental scalars \zthreematII\ and fermions \zthreematIII\
sit in off-diagonal blocks.

The Higgs branch of this gauge theory, along which
$U=V=W$ (as enforced by the quartic scalar interactions inherited from
the ${\cal N}=4$ D-terms) describes motion of the D3-branes
away from the orbifold fixed locus.  The theory also
has a Coulomb branch, where $U=V=W=0$ and components
of the $X^i_k$ for different $k$ get independent VEVs.  This
describes motion of ``fractional'' D-branes away from each
other along the orbifold fixed locus $z^1=0$.

In this case, one finds an effective potential analogous to
that of \TZ\ \tzpot :
\eqn\zthreepot{
\eqalign{
& V_{eff} \sim {g_{YM}^4\over 8\pi^2}{3\over 4}
\Sigma_{a, b=1}^N \biggl[ |x^a_1-x^b_1|^4
log{{|x^a_1-x^b_1|^2}\over{\tilde M^2}} + |x^a_2-x^b_2|^4
log{{|x^a_2-x^b_2|^2}\over{\tilde M^2}}\cr
&+ |x^a_3-x^b_3|^4 log{{|x^a_3-x^b_3|^2}\over{\tilde M^2}}
-|x^a_1-x^b_2|^4
log{{|x^a_1-x^b_2|^2}\over{\tilde M^2}}\cr
&-|x^a_2-x^b_3|^4
log{{|x^a_2-x^b_3|^2}\over{\tilde M^2}}
 -|x^a_3-x^b_1|^4
log{{|x^a_3-x^b_1|^2}\over{\tilde M^2}}
\biggr]\cr
}
}
Here (similarly to the discussion following \tzpot\ )
the first three terms come from integrating out
the four real scalars and the gauge fields which transform in
the adjoint representation, which in this
theory involves 3/4 of the bosons in the theory, hence
the factor of 3/4 relative to the Type 0 result.  The
last three terms come from integrating out the bifundamental
matter, which in this theory consists of one complex
scalar (1/4 of the total bosons) and all of the fermions,
leading to the factor of -3/4 appearing in front of these
terms in \zthreepot.

It is now clear how to generalize this result to arbitrary
$Z_n$ orbifolds.  Consider for example a non-freely acting
$Z_n$ orbifold with rotation angles
$2\pi ({r_1\over n}, {r_2\over n}, 0)$ in the three complex
planes paramaterized by $z^1, z^2, z^3$
transverse to the D3-branes (with $r_1\pm r_2$ even so
that the orbifold acts as a $Z_n$ on all spinors).  
The quiver gauge theory
has a gauge group $SU(N)^n\equiv \Pi_{k=1}^n SU(N)_k$
with one complex scalar corresponding to $Z^3$ transforming
in the adjoint $\Sigma_k {\bf (N^2-1)}_k$.  The complex scalar
corresponding to $Z^1$
transforms in the bifundamental representation
$\Sigma_{k=1}^n({\bf N_k, \bar N_{k+r_1}})$, and that corresponding
to $Z^2$ transforms in the bifundamental representation
$\Sigma_{k=1}^n({\bf N_k, \bar N_{k+r_2}})$.  Half of the fermions
transform in the bifundamental representation
$\Sigma_{k=1}^n({\bf N_k, \bar N_{k+{{r_1+r_2}\over 2}}})$, and the other half
transform in the bifundamental representation
$\Sigma_{k=1}^n({\bf N_k, \bar N_{k+{{r_1-r_2}\over 2}}})$.  From this one obtains
the effective potential
\eqn\genpot{\eqalign{
& V_{eff} = {g_{YM}^4\over 8\pi^2}
\Sigma_{a, b=1}^N \biggl[
{1\over 2}\Sigma_{k=1}^n |x^a_k-x^b_k|^4
log{{|x^a_k-x^b_k|^2}\over{\tilde M^2}}
+{1\over 4}\Sigma_{k=1}^n |x^a_k-x^b_{k+r_1}|^4
log{{|x^a_k-x^b_{k+r_1}|^2}\over{\tilde M^2}}+\cr
&+{1\over 4}\Sigma_{k=1}^n |x^a_k-x^b_{k+r_2}|^4
log{{|x^a_k-x^b_{k+r_2}|^2}\over{\tilde M^2}} \cr
& -{1\over 2}\Sigma_{k=1}^n |x^a_k-x^b_{k+{{r_1+r_2}\over 2}}|^4
log{{|x^a_k-x^b_{k+{{r_1+r_2}\over 2}}|^2}\over{\tilde M^2}}
-{1\over 2}\Sigma_{k=1}^n |x^a_k-x^b_{k+{{r_1-r_2}\over 2}}|^4
log{{|x^a_k-x^b_{k+{{r_1-r_2}\over 2}}|^2}\over{\tilde M^2}}
\biggr]\cr
}
}

\medskip

Finally we note that for orbifolds which act
freely on the $S^5$, with nontrivial
rotation angles $2\pi (r_1/n, r_2/n, r_3/n)$ on the
coordinates $z^1, z^2, z^3$, there are no adjoint scalars
and no Coulomb branch.  This follows geometrically
from the fact that the D-branes span the same dimensions
as the orbifold plane and cannot move apart into separate
fractional branes at the fixed point.  We will return to this in \S3.

\newsec{Tachyon Condensation and QCD}

We have seen in the above section that the quiver theories corresponding
to orbifolds with fixed points on the $S^5$ develop a Coleman-Weinberg
potential on the classical moduli space at one loop,
and we will see in this section that there are interesting unstable directions
in which twisted operators get VEVs.

\subsec{Counting Powers of N}

Let us first clarify and interpret in terms of
the gravity side the
$N$-dependence of the results \tzpot\zthreepot\genpot.\foot{The results on $N$-dependence
here, some aspects of which appear in \TZ, were developed in discussions with O. Aharony.}
Let us first determine the N-dependence of the 1-loop
potential term in the field theory.  In all
of the orbifold theory potentials we have derived, as noted for the Type 0 case in \TZ, the
coefficient of the logarithmically divergent 4-point interaction
among the scalars contains no powers of $N$ beyond that
in the factor of $g_{YM}^4$ after terms of the
form $g_{YM}^4N\Sigma_a|x_a|^4$ cancel out of $V_{eff}$,
leaving terms proportional to $g_{YM}^4(Tr X_1^i X_1^j- Tr X_2^i X_2^j)^2$
and $g_{YM}^4(Tr X_1^2- Tr X_2^2)^2$ at the level of four-point graphs.
Let us rescale the $X$'s so that a factor of
$1/g_{YM}^2=N/\lambda_{'t Hooft}$ appears
multiplying the whole tree-level action.  Then $g_{YM}^2$ counts
loops, and the one-loop potential scales like $g_{YM}^0=1$, down
by a factor of $g_{YM}^2 \sim 1/N$ from tree level.

With this normalization of the fields,
correlation functions of the single-trace operator $N tr X^2$
scale like $N^2$ plus terms
subleading in $1/N^2$.  These correspond to connected genus-zero amplitudes
involving single-particle states on the gravity side \juan\adscft.

Normalizing operators of the form $(tr X^2)^2$ with a power
of $N^2$:
\eqn\Nnorm{
{\cal O}_{double-trace}=N^2(tr X^2)^2,
}
we obtain $l$-point correlation functions of the ${\cal O}_{double-trace}$
which scale like $N^{2l}$ in the free theory, corresponding to 
$l$ disconnected genus zero diagrams describing $l$ strings propagating
across the AdS.  This is in line with the interpretation of multitrace
operators as multiparticle states on the gravity side in the unperturbed theory.  

As discussed below \typezyuk, the twisted operators in our theory
are of the form $tr X_k^2-tr X_{k^\prime}^2$.  Because they transform non-trivially under
the quantum symmetry of the orbifold, terms in the
Lagrangian linear in these operators are not
generated dynamically, but terms quadratic in these operators
which are invariant under the quantum symmetry are  (they are
implicit in the potentials \tzpot\zthreepot\genpot\
calculated out along the Coulomb branches in the last section).
These are double-trace operators, which are thought to correspond
to multiparticle excitations of the dual gravity theory \juan\adscft.
As just discussed, as they appear at one-loop these contributions
scale like
\eqn\Nac{
\delta S \sim \int (tr X^2)^2.
}


Now consider adding a
contribution of the order $(tr X^2)^2$ to the action 
(as occurs dynamically in our theory \Nac ).  Bringing
down a power of \Nac\ into correlation functions, one finds
the leading-N effect from factorized terms of the form
\eqn\factorized{
\langle :(Tr X_k^2- Tr X_{k^\prime}^2)^2:{\cal O}_1,\dots, {\cal O}_l\rangle
\sim \langle (Tr X_k^2-Tr X_{k^\prime}^2 ){\cal O}_1,\dots, 
{\cal O}_{l^\prime}\rangle 
\langle (Tr X_k^2-Tr X_{k^\prime}^2) {\cal O}_{l^\prime +1},\dots, 
{\cal O}_{l}\rangle 
}
where we have replaced $(Tr X^2)^2$ in \Nac\ with the more
precise form $(Tr X_k^2- Tr X_{k^\prime}^2)^2$ we have for deformations
of our theories.  
These go like $N^2$.  However if all the $ {\cal O}_1,\dots, {\cal O}_k$
are {\it untwisted} operators, then each factor in the factorized leading-N
contribution vanishes, and one is left with an effect that is down
by $1/N^2$ from genus-zero effects.  This is in accord with large-N
inheritance on the gravity side \ksorb\ and the field theory
side \bkv, which ensures that at large N the correlators of
untwisted operators are the same as in the ${\cal N}=4$ theory.
The twisted operators do not exist in the parent theory, and
are not constrained by inheritance.

It is interesting that a mass scale $\tilde M$ appears in correlators
of twisted operators at genus zero.  In particular, the couplings
of the double trace
operators $\Sigma_k(Tr X_k^2- Tr X_{k^\prime}^2)^2$ have
nontrivial beta functions at one-loop (as can be seen from
the four-point function contribution to the effective
potentials calculated in \S2).  
So even before we go out on the Coulomb branch, 
the theory is nonconformal 
at leading order in N 
in a nontrivial regime of $\lambda_{'t Hooft}$.  
This is invisible to the untwisted operators alone
at this order in N, in accord with \bkv.  Even so, this is puzzling because of the 
general arguments
advanced in \ksorb\ for the large-N conformality of these theories.
In the next subsection, we will discuss this puzzle and several possible
resolutions which it will be interesting to pursue once we have
pushed the relevant technology to the necessary level.

\subsec{Orbifolding and Symmetries: A Puzzle}

As discussed in \ksorb, there is a fairly general reason to believe that
orbifold field theories should have conformal invariance at large
N, including the physics of twisted operators.\foot{We thank T. Banks and
S. Kachru for discussions of this.}  The worldsheet sigma model
describing strings propagating on the parent space $AdS_5\times S^5$
has a symmetry corresponding to the $SO(4,2)$ isometries of the
$AdS_5$, which commutes with the $SO(6)$ of the $S^5$ and
in particular commutes with an action of $\Gamma\subset SO(6)$ on $S^5$.  
This symmetry commutes with the Hamiltonian of the worldsheet
theory, and therefore all of its correlation functions respect it.
This parent sigma model has many operators, some subset of which
$\{ V_{parent} \} $ constitute mutually local dimension (1,1) vertex
operators describing physical string states.  
When we orbifold, for example by the $Z_n$ actions
we are considering in this paper, we include only those
vertex operators $\{V_{untwisted}\} \equiv (\{ V_{invariant} \}\subset  \{ V_{parent} \})$
which are invariant under the orbifold group action.  Having done
this one can (and should at the one-loop level) add ``twisted'' operators
which are further operators from the set of operators in the parent
sigma model which are mutually local with respect to the reduced set of
operators $\{V_{untwisted}\}$.  So finally 
$\{ V_{orbifold}\} = \{V_{untwisted} + V_{twisted} \} $ gives the
full set of vertex operators for the orbifold theory.  The Hamiltonian
of the full worldsheet sigma model is the same in all of these
theories, and commutes with the $SO(4,2)$.  So the orbifold
theory should have this symmetry and the QFT dual to it
by AdS/CFT should be conformally invariant for all $\lambda_{'t Hooft}$
at leading order in the $1/N^2$ expansion.  This argument appears
rather general (though unforeseen subtleties involving
RR fields may render it inapplicable to our case).   

On the other hand, at weak coupling in the quiver gauge theory one
finds (as we have discussed) nontrivial beta functions for the 
double-trace quartic scalar interactions of the form 
$\Sigma_{k, k^\prime}\lambda(Tr X_k^2- Tr X_{k^\prime}^2)^2$.  Although
it is made out of twisted operators, this contribution to
the Lagrangian does not
itself transform under the quantum symmetry and if we do not
condense the twisted operators we should not have left the 
orbifold point.  

We do not yet know the resolution of this puzzle, but can see
several interesting possibilities (which are not
all mutually exclusive):

\noindent (1)  The above argument about the symmetries
is correct and applies to the RR sigma models of interest
here.  This would suggest that there is a line of
fixed points corresponding to the radius of $AdS_5\times S^5/Z_n$.
Since starting at weak coupling on the field
theory side there is not such a fixed line, this line
of fixed points would have to be fundamentally strongly 
coupled.  

Then the theories we consider here, with running
$\lambda_{ab}^{ij}$, are deformations away from the line
of CFTs dual to the standard orbifold of $AdS_5\times S^5$.  
But these theories share many properties with the standard
orbifold, in particular the quantum symmetry and the inheritance
of untwisted operator correlation functions at large N.  Therefore
even if (1) is true we feel it is important to understand the
gravity-side description of our (perhaps nonstandard) orbifold models.   
This leads to possibility

\smallskip

\noindent (2)  The double-trace operators in the 
effective Lagrangian
of our models correspond to a novel type of worldsheet string theory
on the gravity side, such as the one under investigation 
independently in a supersymmetric context with marginal double-trace
perturbations \ofer.\foot{We thank O. Aharony and M. Berkooz
for sharing with us their ideas on this.}  This novel string theory,
if it exists and applies to our models here, may not have all
the properties required for the above symmetry argument.  As discussed
above, a relative of this possibility is the possibility that
RR sigma models do not satisfy the assumptions in the symmetry
argument presented above.  

\smallskip

Finally, there is always the possibility

\noindent (3)  Phase transitions and/or other unconstrained non-supersymmetric
dynamics ruin the application of the duality to this non-supersymmetric
context.  Because we began with a parent system with two dual descriptions,
the procedure applied to one of them producing the orbifold theory ought
to have a translation into the dual variables if it exists nonperturbatively.
This translation to the dual may not be a standard orbifold construction,
however, which may relate to point (2).  Indeed a phase transition in this
type of system is suggested by the large-radius duality map, which maps the
tachyon to a complex-dimension operator in the field theory \Igortach.  
It would be very interesting to understand better what this means for the duality,
but in this paper we will continue to focus on the small-radius (weak
't Hooft coupling) regime.      

\smallskip

Because of the RR fields and strong coupling
issues, establishing the precise resolution of this
puzzle appears out of reach of current
technology, and we will leave it for future work.
We think it is likely that there is a resolution (perhaps
along the lines of (1) and/or (2)) which preserves the duality and
teaches us something new about the gravity side, and we will proceed
with our analysis on the assumption that the duality holds.
In particular, our analysis 
has generated further concrete evidence
in favor of the duality (in addition to generating the puzzle
discussed in this subsection).  
However, possibility (3) should be kept in mind.

\subsec{Tachyons and AdS/CFT Duality}

It has been suggested \typezero\Igortach\ that the Type 0 tachyon
is lifted at small $AdS$ radius to satisfy
the Breitenlohner-Freedman bound \BF.  Heuristically this
might be expected from the fact that the $AdS$
curvature reaches string scale for small enough 't Hooft
coupling, so that a string-scale tachyon need not violate
the bound \typezero.  This sort of behavior has been
seen in the $AdS_3$ context in \hirosijuan.  
Further, the twisted operator $tr F_1^2+DX_1DX_1-F_2^2-DX_2DX_2$,
to which the tachyon couples directly at large radius,
is actually slightly irrelevant
at weak coupling \Igortach, which according to 
an uncorrected extrapolation of the large radius
duality map would translate to a non-tachyonic mass in the
bulk gravity theory.

However, we have seen that the weakly coupled dual field theory
has instabilities in the potential at leading order in $1/N^2$
which cause certain twisted operators (which
have the same discrete quantum numbers as the
large-radius tachyons) to condense, either directly
or via tunneling depending on the small-$X$ behavior
of the potential.
That the instability
appears at genus zero shows that it persists even in the strict
large-N limit.  This demonstrates an instability of
the string theory which causes modes from the orbifold twisted
sectors to condense even at small radius.

As we have discussed, because of the running
couplings in the theory, our one-loop analysis is not sufficient
to determine whether the instability is perturbative or requires
a non-perturbative tunneling process to access.  
If it is non-perturbative, the situation is reminiscent of those described in
\tuninst, where a tachyonic instability in one limit of
moduli space appears to turn into a non-perturbative instability
mediated by a gravitational instanton in another limit.

\subsec{Patterns of Symmetry Breaking}

In the remainder of this section we will provide
a preliminary discussion of the physics that results when the
twisted operator VEVs
turn on.  We will begin with some heuristic intuition from the gravity side,
and then analyze concretely some aspects of the Higgs structure of the model given the
scalar potentials calculated in the previous section.

On the gravity side, we expect perturbative tachyon condensation to produce
a subcritical dimension target spacetime \joetach\anttach.  This
is because the zero-momentum tachyon vertex operator is a relevant
operator on the string worldsheet.  The worldsheet beta function
equations are then satisfied by a nontrivial field configuration
for the dilaton, metric, and other string fields; in particular
dilaton gradients contribute effective central charge to compensate
for that lost by going to a subcritical dimension (as occurs
for example in the case of a linear dilaton with flat string-frame
metric) \joetach.  In the context of the AdS/CFT correspondence,
the dimensions of the $S^5/Z_n$ and the radial direction
of AdS arose from the directions transverse
to the D3-branes, which are parameterized by worldvolume
scalars.  It is natural to expect therefore that losing 
and/or deforming the
$S^5$ and radial dimensions would correspond to losing the scalars in the
dual quiver gauge theory, and perhaps also the fermions which
also transform under $SO(6)$ rotations.

In situations with non-perturbative instabilities on the gravity side
\tuninst\ one also has a sense in which degrees of freedom
are lost, as one ``tunnels to nothing''.\foot{From other points
of view one appears to tunnel to flat space via a Schwinger
effect \gary\gutperle, a situation whose interpretation and
whose relation to our results here would be very interesting to
clarify.}  In our situation, as we will discuss shortly, one
does not always expect to decay to nothing, but one can decay
to something which is in some sense less
than what one had to begin with:  from the full quiver gauge theory
to a long-distance sector with pure glue QCD.

Let us discuss some patterns of symmetry breaking that emerge
from our potentials \tzpot\zthreepot\genpot.
There are instabilities
in the effective potential correponding to VEVs for twisted
operators in the gauge theory, manifested in the D-brane language
convenient for the calculations in \S2\ by relative motion of fractional
branes along the orbifold plane described by VEVs for diagonal entries
of the adjoint scalar matrices.  Let us investigate the effect of
turning on these VEVs.

Let us analyze the Type 0 case \tzpot\ for simplicity; similar
patterns will emerge in the higher $Z_n$ cases and can be
analyzed in a similar way.
Consider the direction in field space in which $X_1^i$ gets a VEV
\eqn\typezeroVEV{
\langle X_1^i\rangle =diag(\rho_1^i, \rho_2^i, \dots, \rho_{N-1}^i,
-\rho_1^i-\rho_2^i-\dots-\rho_{N-1}^i),
}
which satisfies the $SU(N)$ condition
\eqn\sun{\Sigma_{a=1}^N \rho_a =0,
} 
and in which $\langle X_2^i\rangle = 0$.
  
Plugging this into the effective potential \tzpot, we obtain
\eqn\potrho{
V(\vec{\rho_a})\sim g_{YM}^4\biggl(
\Sigma_{a, b}|\vec\rho_a - \vec\rho_b|^4 log {{|\vec\rho_a - \vec\rho_b|^2}\over
{\tilde M^2}}-2N \Sigma_a |\vec\rho_a|^4 log {{|\vec\rho_a|^2}\over
{\tilde M^2}}\biggr)
}
where we have replaced the transverse $R^6$ index $i$ by vector
notation.  

Let us consider the minimization of this potential with
respect to the $\rho_a^i$.  The first term in \potrho\ describes the 
force between ``electric'' branes (those with $SU(N)_1$ on their 
worldvolume).  The second term describes the force between the electric branes
and the ``magnetic'' branes (those with $SU(N)_2$ on their worldvolume)
which are all sitting at the origin, $\vec X_2=0$.  
At sufficiently small distances between the 
branes, the former is repulsive and the      
latter is attractive.  

There is a relatively simple configuration where the $\vec \rho_a$ are arranged
symmetrically (equally spaced) 
on an $S^5$ of radius $\rho$.  This satisfies the $SU(N)$
constraint \sun.   
It is an extremum of the effective action in the angular
directions.  By playing the mutual repulsion of the electric
branes against their attraction to the magnetic branes at
the center, we
will find a minimum for the radial mode $\rho$, 
generalizing the one along the direction
$\langle X_1^i\rangle =diag(\rho, 0, \dots, 0, -\rho)$ discussed in \TZ.

Approximating the sum over branes indexed by $a$ by an integral over the
angles of the $S^5$, we obtain an approximate form of the potential
which will be sufficient to indicate the presence of the anticipated minimum:
\eqn\potint{
{V\over g_{YM}^4}\sim \int d\Omega_1 d\Omega_2 |\vec\rho(\Omega_1)-\vec\rho(\Omega_2)|^4
log {{|\vec\rho(\Omega_1)-\vec\rho(\Omega_2)|^2}\over{\tilde M^2}}
-2N \int d\Omega_1 |\vec\rho(\Omega_1)|^4 log 
{{|\vec\rho(\Omega_1)|^2}\over{\tilde M^2}}
}
Using this, separating the radius $\rho$ of the sphere of fractional D-branes
from the angular variables, we obtain
\eqn\potrad{
V(\rho)\sim N^2 g_{YM}^4\rho^4 log\bigl({{\rho^2e^{379/240}}\over {\tilde M^2}}\bigr)
}
The potential \potrad\ has a minimum at $\rho$ of order $\tilde M$,
\eqn\minrho{
\rho_{min}^2=\tilde M^2 e^{-1/2}e^{-379/240}
}
as in our earlier discussion of the Coleman-Weinberg potential.
We have therefore balanced the attractive and repulsive forces
as anticipated.
In particular, one finds the force between the electric branes
and the magnetic branes is attractive in this regime.  Because
the forces grow with distance, this suggests that the magnetic
branes at the origin are stable against small fluctuations,
which means the $X^i_{2}$ scalars are massive.  Indeed
this follows from an analysis of small fluctuations in the
$\vec x_{2b}$ directions, as follows. 

Expanding \tzpot\ around the background symmetric 
distribution of $\vec\rho_a$, we find
the following mass terms for the $x_{2,b}^i$:
\eqn\xtwomasses{
-2g_{YM}^4\Sigma_{a,b}\biggl(2|\vec x_{2,b}|^2
|\vec \rho_a|^2 log{{|\vec\rho_a|^2e^{1/2}}\over{\tilde M^2}}
+ 4x_{2,b}^lx_{2,b}^m\rho_a^l\rho_a^m
log{{|\vec\rho_a|^2e^{3/2}}\over{\tilde M^2}}\biggr)
}
Summing over the spherically distributed $\vec\rho_a$, 
the off-diagonal terms in the mass matrix sum to zero and 
we see that the
diagonal terms are nonzero and positive at the minimum \minrho\
(since $log {{\rho^2_{min}e^{1/2}\over \tilde M^2}}=-379/240<0$
and $log {{\rho^2_{min}e^{3/2}\over \tilde M^2}}=1-379/240<0$).\foot{Note that
there is no coupling-constant dependence in
the computation of the signs of forces and scalar $m^2$'s which depend only on
the values of order one numbers arising from the geometry of the configuration
at the minimum \minrho.}

We have now accumulated enough information to determine the effect 
on the gauge theory of turning on this VEV.  It breaks one
$SU(N)$ to $U(1)^{N-1}$ and leaves the other $SU(N)$ intact.
From \typezyuk\ one finds that 
all the fermions get masses
once our VEVs \typezeroVEV\ are turned on.
As we have just seen, the scalars $X_2^i$ get positive mass squared.

The angular fluctuations of the $X^i_1$ on the other hand are
dominated by repulsive interactions between
the electric branes, so these fluctuations appear to be
unstable.  
Once this configuration of twisted VEVs in the gauge theory is turned on,
the long-distance physics of the gauge theory consists of
QCD plus decoupled U(1) factors and unstable scalars.

There is another configuration in which QCD sectors emerge
at low energies without instabilities in the other sectors; this
is a likely endpoint of the spherical configuration we
began with.  It is another natural generalization to large $N$
of the instability discussed in \TZ.  Consider again
$\langle \vec X_2=0\rangle $.  Take 
$\langle X_1 \rangle = diag(\vec r, \vec r, \dots, \vec r, -\vec r,
\dots, -\vec r)$ where the first $N/2$ diagonal entries are $\vec r$
and the last $N/2$ entries are $-\vec r$.  Geometrically, this
describes $N/2$ electric branes at $-\vec r$ and $N/2$ electric
branes at $+\vec r$, with $N$ magnetic branes at the origin.

In this case, there is a minimum at
\eqn\minrholine{
{{ r_{min}^2e^{1/2}}\over {\tilde M^2}}=2^{-8/3}
}    
about which the fluctuations of both $\vec X_1$ and $\vec X_2$ work
out to be massive, which again is consistent with
naive expectation from the signs of the forces in the vicinity
of the minimum \minrholine.  The Yukawa couplings \typezyuk\
yield masses for all the bifundamental fermions in this configuration.

This more stable configuration leaves, on the
gauge theory side at distances
long compared to the masses of the scalars and fermions, 
a pure glue QCD sector with gauge
group $SU(N)$, decoupled from two others with gauge
groups $SU(N/2)$ and a relative $U(1)$ factor.    

Because of the limited range of validity of the 1-loop
calculation of $V_{eff}$, we do not know if the potential 
is bounded or unbounded from below at large $\langle X\rangle $.
If it turns out to be unbounded, it is tempting to 
suggest that for infinitely large VEVs for the twisted
operators in the gauge theory, the theory may reduce to separate
pure glue QCD plus decoupled $U(1)$ sectors.  However because of
the long-range forces on this fractional D-brane branch
of the gauge theory, it is not clear what the masses of
the $X_2$'s will be as a function of the VEVs of the $X_1$'s,
and it is logically possible that the $X_2$ scalars would
come back down to zero mass and/or become unstable
as $X_1$ increases beyond the regime of validity of our
current calculation.    

In our $Z_n$ orbifolds, before going out on the Coulomb branch the
gauge group is $SU(N)^n\equiv \Pi_{k=1}^n SU(N)_k$.
Turning on VEVs for the diagonal elements of the $X$'s similarly leads
to the near-horizon limit of various 
fractional D-brane configurations whose low-energy theories
involve gauge symmetry breaking and some massive and/or decoupled scalar
and fermion matter.  It would be interesting to classify all the
possible behaviors in arbitrary $Z_n$ orbifold models based on
the potentials derived in the last section, but we will leave
that for future work.    

It is not clear from this
analysis whether this M-theory dual
to the remaining gauge theory will be a perturbative string theory.\foot{The
results \kutsei\ perhaps suggest otherwise.}  Indeed
in ordinary large-radius string backgrounds the tachyon
mixes with the dilaton, and its condensation leads to strong
dilaton gradients; the analogous phenomenon should
be expected in our small-radius AdS/CFT system.  (Correspondingly
on the gauge theory side, the VEVs for twisted operators that
we have turned on can induce large renormalizations of all
the couplings in the gauge theory.)  
Even before turning on VEVs for twisted operators along the
unstable directions, a novel kind of string theory
may be required on the gravity side of a dual pair
in which the field theory is deformed by a double-trace
operator of this kind, as discussed in \S3.2 \ofer.   

In any case, we have arrived at
an interesting answer via AdS/CFT duality to the question of
what can happen when one condenses a tachyon in closed
string theory:  in this system, it rolls to the
gravity dual of a gauge theory with less symmetry and reduced
matter content, but sometimes retaining pure glue QCD
factors in the infrared.

As discussed above, condensing the tachyon is expected
to lead to a subcritical matter sector on the string worldsheet,
and we have just learned that the corresponding process
on the gauge theory side can lift or decouple the extra matter and gauge
fields beyond pure glue QCD.
Noncritical string theory was conjectured
to be dual to QCD in \polyakov.  Our results provide
some further evidence in this direction.

\newsec{Freely Acting Orbifolds and Tachyon-freedom}

Consider an orbifold group $\Gamma\subset SO(6)$ which fixes
an isolated point in $R^6$.  In the presence of $N$ D3-branes centered
at the fixed point, the spacetime geometry blows up
into a near-horizon region which is completely smooth, with the
orbifold acting freely on the $S^5$.

Because the orbifold fixes an isolated point in $R^6$, the codimension of
the singularity is the same as the codimension of the D3-branes,
so the spacetime has no directions along which
fractional branes could separate. Correspondingly, the
scalars in the resulting quiver gauge theory are all
in bifundamental representations, in contrast to the above
non-freely-acting cases where the scalars describing motion
along the orbifold fixed locus remained in the adjoint.  Thus, for
freely acting orbifolds, there is no Coulomb branch.

Since in these cases the classical moduli space does not
include a branch where twisted states can develop
a VEV, the theory will remain stable to all orders in $\lambda_{'t Hooft}$.
We therefore suspect
that there are no twisted instabilities for any radius (any
't Hooft coupling $\lambda_{'tHooft}$)
in this system, though there is a logical possibility
that one develops at a $\lambda_{'tHooft}$ of order one.
(If so, it would have to disappear again for large $\lambda_{'tHooft}$
as discussed above.  This result is a new 
prediction for the gravity description
at small radius.)\foot{Very naive gravity-side intuition might
have suggested that a tachyon would arise at small radius,
since the positive mass squared of twisted states
at large radius is driven
by their winding energy around the $S^5/\Gamma$, and when the $S^5/\Gamma$ 
becomes small the winding energy would appear to be negligible.
However, the substringy dynamics of curved Ramond backgrounds
is hardly a place where naive intuition applies.  Our QFT
analysis of the lack of instability in this system at
small radius is a prediction for the worldsheet RR sigma model.}

\newsec{$Z_n\times Z_n$ Orbifolds with Discrete Torsion}

In more general orbifolds than those we have considered so
far, such as $Z_n\times Z_n$ orbifolds, one can project out
the lowest-lying twisted-sector tachyons with a nontrivial
choice of discrete torsion.  However, at large radius, tachyonic
modes dressed with angular momentum along the
$S^5/\Gamma$ will survive this projection.  It is interesting to consider
whether this instability persists at small radius (weak
't Hooft coupling) in these theories.

In this section we will study in particular
a $(C/Z_3)\times (C/Z_3)$ orbifold with nontrivial discrete torsion.
Let the first $Z_3$, generated by $g_1$,
act on the $z^1$ direction, and the second generated by $g_2$
on the $z^2$ direction, with the third complex plane parameterized
by $z^3$ left invariant.  The $g_1$ twisted sector vacuum
energy is -1/3, as is that of the $g_2$ twisted sector, and
in the absence of nontrivial discrete torsion these states
correspond to physical tachyons in spacetime.
A nontrivial choice of discrete torsion projects out each of
these vacua ($g_1$ projecting out the vacuum in the $g_2$ twisted
sector and vice versa).  However, there are momentum states
invariant under the both $g_1$ and $g_2$ which still have
tachyonic masses $m^2<0$ in spacetime.

Naive intuition might suggest that these momentum-mode tachyons
may get lifted as we go toward small radius since the momentum
contribution to the $m^2$ of the state grows as the radius shrinks.
Again naive intuition is liable to fail in these highly curved
Ramond backgrounds, and as in the previous examples, the QFT instability analysis is the
appropriate method for answering this question at small
radius given the limitations of current technology on
the gravity side.

The worldvolume theory of D-branes on orbifolds with discrete
torsion was worked out in \DT\ (and studied in the context
of AdS/CFT in \berleigh).
The result for our case in particular is as follows.
The theory is an $SU(3N)$ gauge theory
with three complex scalars $Z^{1,2,3}$ and four
Weyl spinors $\chi^{1,\dots, 4}$
in the adjoint, i.e. a theory with the field content of
${\cal N}=4$ $SU(3N)$ SYM.  The interactions, however, differ
from those of the ${\cal N}=4$ theory.  The quartic scalar
couplings involving $Z^1$ and $Z^2$ are deformed from the
usual commutators to take the form:
\eqn\Lscal{\eqalign{
& {\cal L}_{scalar} = tr \biggl[(Z^1Z^2-\alpha Z^2 Z^1)
(Z_{\bar 2}Z_{\bar 1}-\alpha^{-1}Z_{\bar 1}Z_{\bar 2}) \biggr]\cr
& tr \biggl([Z^1,Z^3][Z^1,Z^3]^\dagger\biggr)
+ tr \biggl([Z^2,Z^3][Z^2,Z^3]^\dagger\biggr)\cr
}
}
where $\alpha=e^{2\pi i/3}$.  Similarly the Yukawa couplings are
deformed to the form
\eqn\yukDT{\eqalign{
& \alpha^{-1}\Gamma_1^{\alpha\beta}
tr \biggl[\chi^\alpha(Z^1\chi^\beta-\alpha^{-1}\chi^\beta
Z^1)\biggr]
+ \alpha\Gamma_2^{\alpha\beta}
tr \biggl[\chi^\alpha(Z^2\chi^\beta-\alpha\chi^\beta
Z^2)\biggr]\cr
& + \Gamma_3^{\alpha\beta}
tr\biggl(\chi^\alpha [Z^3, \chi^{\beta}]\biggr)\cr
}
}

The Coulomb branch describing fractional branes is parameterized
by diagonal $Z^3$ matrices.  We can now immediately observe
a difference between this case and the cases discussed in \S2.
Namely, the one-loop Coleman-Weinberg potential will be absent
here, since all the tree-level vertices involving $Z^3$ are exactly the
same as in an $SU(3N)$ ${\cal N}=4$ theory.  On the other hand,
higher-loop contributions to the effective potential of the
gauge theory mix $Z^3$ with all the other fields, and we expect
such contributions will get generated.  It would be interesting
to explore their signs to see if an instability exists
in this case at higher orders in $\lambda_{'t Hooft}$.

\newsec{Discussion and Future Directions}

In this paper we have identified global instabilities in certain weakly-coupled
non-supersymmetric gauge theories whose AdS/CFT duals contain twisted-sector
tachyons at large radius.  These instabilites, which induce
VEVs for twisted field theory operators, appear at one-loop in the gauge theory and
genus zero in the string theory, though their effect on untwisted operators is
suppressed by factors of ($1/N^2$), as expected from large-N inheritance.

At higher orders in $1/N^2$ there will be a rich set of
dynamically generated contributions to the effective action which are not
constrained by large-N inheritance from the parent ${\cal N}=4$ theory.  This can
(and presumably does) include quadratically divergent scalar masses (as well
as quartically divergent vacuum energy which does not affect the QFT dynamics).
It would be interesting to calculate these effects and understand their description
on the gravity side of the correspondence.  These finite-N effects
can have a dramatic effect on the matter content and dynamics,
and it is necessary to calculate these in order to understand the finite-N
system.  Some interesting
perturbative calculations in these theories
were done for example in \othercalcs.  While we feel such an analysis
is further motivated by our work here, it is somewhat subtle to carry out since the
QFT couplings appropriate to the gravity dual may themselves be shifted from the
orbifold values by contributions of order $1/N^2$.\foot{We thank M. Strassler for
reminding us of this difficulty.}

As discussed in \S2, it would be very interesting to 
ascertain the behavior of the effective potential
near the origin of the Coulomb branch.  It would also be interesting
to see whether a higher-loop analysis leads to 
persistant instability at large $\langle X\rangle $, and
to study the meaning of (and possible constrints on) the quiver theory
renormalization conditions from the gravitational dual.
These last issues mirror the difficulty on the gravity side
of determining the form of the tachyon potential when the
tachyon VEV is large and mixes strongly with other string
fields.

One important generalization to consider is a case where
quark flavors survive the tachyon condensation process, so
that we get more than just the pure glue QCD theory.  Recall
that in our tachyon condensation process in \S3\ the fermions
decoupled and/or became massive as the twisted operator's
VEV turned on.
A case where flavors survive
may well involve a second set of D-branes in addition
to the D3-branes contributing the $SU(N)$ gauge group, so
that the gravity side has open strings as well as closed strings.
It would be interesting to perform a general analysis
of symmetry breaking patterns for the $Z_n$ orbifolds
considered here and more general ones; the configurations
we discussed in \S3.4\ are particularly simple and
there may be a rich set of interesting possibilities
implicit in the potentials \tzpot\zthreepot\genpot.  

It would be interesting to explore the tachyon potential in
closed string field theory in this system, and compare the
Coleman-Weinberg
potential to the closed-string analogues of formulas
in \opentach\ for
the tachyon potential in open string field theory.  This
is out of range of current technology, and our QFT calculations
are simply predictions for the behavior of the appropriate
gravity side sigma model.

It would also be interesting
to study the tachyon potential on the gravity side at
large radius, to see what happens to the $S^5/\Gamma$
and to the RR fields upon
tachyon condensation in that regime.  We may be able to
get a handle on this by studying the geometry of fractional
D-branes splitting apart, via nonsuper-gravity at
large radius.  It would also
be very interesting to explore potential relations to
the work of \maldstrom\ and \gutperle.  With respect
to the latter, one would need to repeat our analysis
of D3-branes in the Type 0B theory
for the (nonconformal) even-dimensional D-branes of the Type 0A
theory, where the conjectured dualities and instabilities of \gutperle\
might apply most directly.

In general, it is important to improve our understanding of the
gravity side of the duality (and the duality map) in order to resolve the 
puzzle of the violation of conformal invariance on the field
theory side discussed at length in \S3.2.  

Finally, it would be interesting to study further examples
of tachyonic and non-tachyonic non-supersymmetric AdS/CFT duals, to see
how general the
pattern found here of large radius instabilities persisting
to small radius proves.  We covered a large
class of examples in our analysis here, but there are many
more cases that could be considered.

\medskip

\noindent{\bf Acknowledgements}

We would like to thank O. Aharony for very useful comments and
questions on this project 
and for sharing with us his insights with M. Berkooz on double-trace operators
in an independent context \ofer.  
We would also like to thank T. Banks, M. Berkooz, O. DeWolfe, D. Gross, A. Hashimoto, G. Horowitz,
S. Kachru, I. Klebanov, D. Kutasov, A. Lawrence, H.  Ooguri, 
S. Sethi, S. Shenker, M. Strassler, A. Strominger,
J. Troost and K. Zarembo for interesting and helpful discussions on this and/or related
topics.  We thank the participants of the ITP M-theory
workshop for their questions and comments during a discussion
of an earlier incarnation of this work. Our research is supported in
part by the Department of
Energy under contract DE-AC03-76SF00098, by NSF grant PHY-95-14797, by
A DOE OJI grant, and by an Alfred P. Sloan Foundation Fellowship.
We are grateful for the hospitality of the Institute for Theoretical
Physics at UCSB, where we receive support from the NSF under grant
number PHY-99-07949.  The research of A.A. was also supported in
part by an NSF Graduate Fellowship.

\listrefs

\end